\documentstyle[hip-artc]{article}  
\volnumber{15}  \edyear{2002} \frompage{000} \topage{000}                
\recrevdate{15 November 2002}                                                
\def\rightmark{Wigner Functions in High Energy Physics} 
\def\leftmark{T. Cs\"org\H{o}}
\newcommand{\fsp}[2]{\epsfysize=#1 \epsfbox{#2}}
\title{Wigner Functions in High Energy Physics} 
\authors{\twerm 
{T. Cs\"org\H{o}$^1$ %
\index{{\tt csorgo@sunserv.kfki.hu} .
\rm Dedicated to the memories of  Gy. Marx and E. P. Wigner}
}\\[2.812mm]
{\normalsize
\hspace*{-8pt}$^1$ MTA KFKI RMKI, \\ 
H - 1525 Budapest 114, POBox 49, Hungary\\[0.2ex] 
}}
 
\abstract{Recent developments are (meta)reviewed in the applications of 
Wigner functions to describe the observed single particle spectra and
two-particle Bose-Einstein (or Hanbury Brown -- Twiss) correlations 
in high energy particle and nuclear physics, with examples from 
hadron-proton  and Pb + Pb collisions at CERN SPS.
}
 
\makeindex
\begin{document}
 
\maketitle

\section{Introduction}\label{intro}
High energy physics attempts to determine what are the  most basic constituents
of matter and how these most elementary forms of matter interact 
with each other. The method to achieve this goal is to build 
accelerators at the highest awailable energies. High energy particle physics
studies the collisions of elementary particles and attempts to determine
the interaction between as few particles as possible.
High energy heavy ion physics attempts to create a new medium with the
highest currently available volumes and energy densities so that
the in-medium interactions of the elementary particles could be
determined. These studies can be utilized to learn about the
form of matter in the first few microseconds of the early Universe
as well as about the behaviour of matter in the center of dense (neutron)
stars.

Why are Wigner functions applied in high energy physics?
Wigner functions are  quantum analogies of the classical
phase-space distributions. 
In conventional field theory, e.g. perturbative QCD,
the calculation of the momentum distributions of the observable
multi-particle final state can be performed in momentum space only.
Usually, a description using Wigner functions
can be equivalently rewritten into a calculation in momentum space, too.
In some cases, the properties the phase space description becomes important
in high energy physics. Currently, such important topics are
the search for new states of matter in high energy heavy 
ion collisions using correlation techniques~\cite{panitkin-rev}, 
and the precise determination
 of the mass of the W-boson in $e^+e^-$ annihilation at LEP~\cite{leif}. 
In the first case, a new state of strongly
interacting matter is searched for, where quarks and gluons
are deconfined, not bound into the usual hardonic states. Theoretical
calculations suggest~\cite{fodor-katz} that a new phase of matter can
be reached if the temperature, or, the energy density can be
sufficiently high, ($T \ge 170$) MeV, at the currently reached finite net 
baryon densities. The temperature,  the energy  and the baryon density
are local variables, they depend on the phase space evolution of the
matter (and of course on the applicability of statistical and
thermodynamical concepts in high energy heavy ion collisions).
In case of the $W$-mass reconstruction in $e^+e^-$ collisions at LEP2,
the aim is to decrease the error on the estimate that comes from the
invariant mass estimates of 4-jet decays. However, Bose-Einstein
symmetrization effects may introduce a non-perturbative ``cross-talk"
between the jets\cite{leif}. Hence pions that come from the decay of the $W^+$
may prefer to appear with momenta close to that of another like-charged
pion from the decay of the $W^-$, due to the bosonic nature of the pions.
Calculations suggested that this modification of the momentum distribution
may result is a systematic error as big as 100 MeV in the reconstructed
$W$ mass, which is the biggest systematic error on this observable and
should be reduced to reach the expected level~\cite{kittel-prev}.
However, the most recent paper by the L3 collaboration finds no evidence for
inter-W Bose-Einstein correlation effect in fully hadronic W decays
at LEP2~\cite{L3-interW}.
 
Quantum-statistical correlations are observed
in high energy particle physics in all kind of particle reactions 
ranging from the ``most elementary" $e^+ e^-$ reactions, see e.g.\cite{epem}
through hadron-proton reactions, see e.g.~\cite{hp},
 to the ``most complex" heavy ion reactions, (see ref.~\cite{AA}
for a recent summary of the data),
essentially in all experimentally accessible energy regions.
These quantum-statistical correlations are correlations of 
intensities of particles, and they are attributed to either the 
Bose-Einstein statistics of meson pairs, or, to the Fermi-Dirac statistics
of the baryons. These statistics are dependent, in turn, on the
phase-space densities of particles, so it is not a surprise,
that Wigner functions turn out to be fundamental when one attemts
to describe the experimentally measured two-, three- or higher order
intensity correlations. As this note is very brief and attempts only
to create a narrow bridge between the current quantum optical studies
and the theory and experiment dealing with quantum statistical correlations
and Wigner functions in high energy physics, let us quote some of
 the more detailed recent reviews in the 
field~\cite{AA,kittel-prev,cs-rev,weiner,uli-urs,lorstad,zajc}.
 

\begin{figure}[htb]
\begin{center}
\fbox{\ \fsp{3.6in}{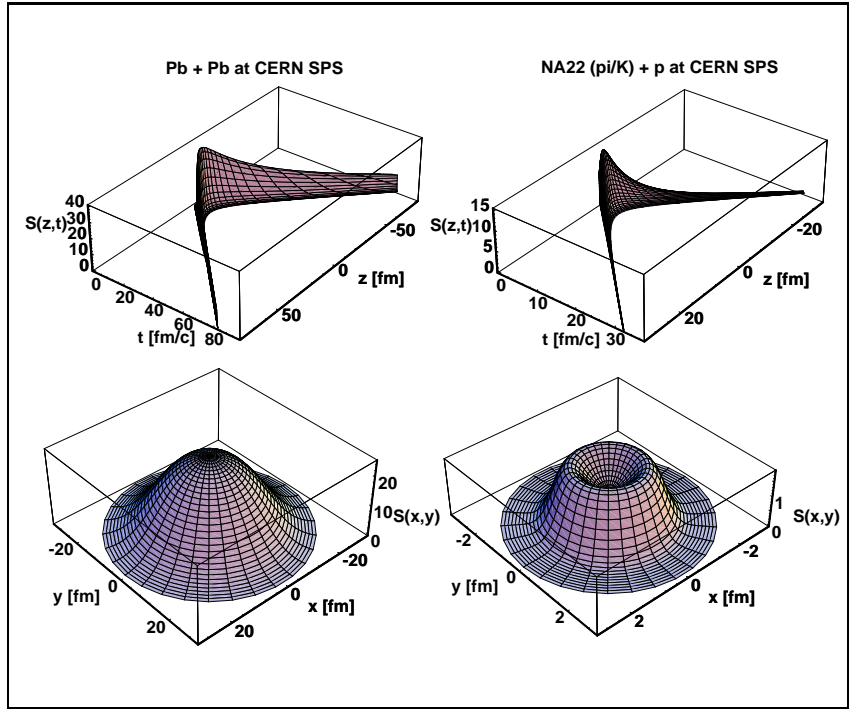} }
\caption[summ-s3.eps]{This figure shows some structures in the Wigner function $S(x,k)$
when integrated over the momentum, and plotted as a function of the $(t,r_z)$ temporal and
longitudinal as well as in the $(r_x, r_y)$ transverse coordinates. 
Till now, non-Gaussian structure of Wigner functions 
has been reconstructed in the transverse distribution
of hadron-proton reactions at CERN SPS energies only, ref.~\cite{cs-rev}.}
\label{summ-s3.eps}
\end{center}
\vspace*{-1.0cm}
\end{figure}

Thus Wigner functions have a broad application in the theory and experiment
of high energy physics, as they provide a natural tool to model these
reactions in terms of phase space distritubutions. 
In particular, they are very popular in describing quantum statistical
correlations (Bose-Einstein or Fermi-Dirac correlations) of the emitted
particles. Currently, Gaussian models dominate~\cite{uli-urs}, and the coupling
between the coordinate space and the momentum space distributions
is frequently modelled with the help of hydrodynamics~\cite{cs-rev}. 
State of art
results attempt to quantify and to characterize the deviations from Gaussian
structure.
Shell of fire type of structures have been reconstructed
in the transverse coordinate distribution in hadron-proton 
reactions~\cite{cs-rev},
and significant non-Gaussian correlation functions were found in $e^+ e^-$ and 
hadron-proton  reactions~\cite{kittel-prev}. Till now, non-Gaussian structure
of the Wigner-function has been reconstructed in $h+p$ reactions 
only,~\cite{cs-rev}. 
The current analysis of correlation data in high energy 
heavy ion physics is confined to the Gaussian approximation, possible
more detailed structures (e.g. binary sources~\cite{binary}) have not
yet been searched for, in this branch. 

The Wigner function formalism has been extended
not only to thermal, or, in the quantum statistical sense, chaotic 
sources, but also to coherent particle emitting sources, for example,
the pion laser model~\cite{plaser}. In discussions of in-medium hadron
mass modification, two-mode squeezed states of hadrons were studied
in terms of Wigner functions, both for bosons~\cite{bbc} and 
for fermions~\cite{fbbc}. In principle, squeezing  
appears in these papers due to sudden disintegration of a medium
where particles propagate with modified mass, so the situation
is very similar to the one considered in quantum optics 
in ref.~\cite{Adam-Janszky}. The only essential difference is that
the field theoretical formulation results in two-mode squeezed states,
hence quanta with opposite momenta and opposite quantum numbers will
be correlated. Wigner functions appear in the formalism because the
strenght and the width of these back-to-back correlations
depends on the phase-space distribution of quanta in the
mass-modified medium~\cite{bbc,fbbc}.

From a broader perspective, it seems that Wigner functions will be 
indispensable tools in high energy physics as long as 
effects sensitive to the phase-space evolution of these reactions
will be investigated.

This work was partially
supported by an 
by OTKA grants T038406, T034269 and T029158 
and by the {MTA-OTKA-NSF} grant {INT0098462}.
 
\begin{notes}
\item[a]
E-mail: csorgo@sunserv.kfki.hu
\item[b]
Dedicated to the memories  of Gy. Marx and E. P. Wigner.
\end{notes}

\vfill\eject
\end{document}